\documentclass[sigconf]{acmart}

\AtBeginDocument{%
  \providecommand\BibTeX{{%
    \normalfont B\kern-0.5em{\scshape i\kern-0.25em b}\kern-0.8em\TeX}}}

\copyrightyear{2022}
\acmYear{2022}
\setcopyright{acmlicensed}
\acmConference[CPSS '22]{Proceedings of the 8th ACM Cyber-Physical System Security Workshop}{May 30, 2022}{Nagasaki, Japan}
\acmBooktitle{Proceedings of the 8th ACM Cyber-Physical System Security Workshop (CPSS '22), May 30, 2022, Nagasaki, Japan}
\acmPrice{15.00}
\acmDOI{10.1145/3494107.3522773}
\acmISBN{978-1-4503-9176-4/22/05}

\usepackage{xcolor}
\usepackage[detect-weight=true,detect-family=true,group-separator={,}]{siunitx}
\usepackage{xspace}
\usepackage[inline]{enumitem}
\usepackage{nicefrac}
\usepackage{booktabs}

\newcommand{\ie}{\textit{i.e.},\xspace}
\newcommand{\eg}{\textit{e.g.},\xspace}
\newcommand{\cf}{\textit{cf.}\xspace}

\newcommand{\wrt}{w.r.t.\xspace}

\newif\ifdraft

\drafttrue
\draftfalse

\ifdraft
\newcommand{\todo}[1]{{\color{red}{\textbf{TODO:} #1}}}

\newcommand{\CUT}[1]{{}}
\newcommand{\mh}[1]{{\color{orange}{#1}}}
\newcommand{\mhtodo}[1]{{\color{orange}{\textbf{TODO:} #1}}}
\newcommand{\ew}[1]{{\color{purple}{#1}}}
\newcommand{\ewtodo}[1]{{\color{purple}{\textbf{TODO: #1}}}}
\newcommand{\jp}[1]{{\color{green}{#1}}}
\newcommand{\kw}[1]{{\color{lightblue}{#1}}}
\newcommand{\kwtodo}[1]{{\color{lightblue}{\textbf{TODO:} #1}}}
\newcommand{\dk}[1]{{\color{olive}{#1}}}
\newcommand{\dktodo}[1]{{\color{olive}{\textbf{TODO:} #1}}}
\newcommand{\md}[1]{{\color{blue}{#1}}}
\newcommand{\ina}[1]{{\color{violet}{#1}}}

\else
\newcommand{\todo}[1]{}

\newcommand{\CUT}[1]{{#1}}
\newcommand{\mh}[1]{{}}
\newcommand{\mhtodo}[1]{{}}
\newcommand{\ew}[1]{{}}
\newcommand{\ewtodo}[1]{{}}
\newcommand{\jp}[1]{{}}
\newcommand{\kw}[1]{{}}
\newcommand{\kwtodo}[1]{{}}
\newcommand{\dk}[1]{{}}
\newcommand{\dktodo}[1]{{}}
\newcommand{\md}[1]{{}}
\newcommand{\ina}[1]{{}}
\fi

\newcommand{\step}[2][]{Step~\raisebox{.5pt}{\textcircled{\raisebox{-.5pt}{\small#2\raisebox{0.5pt}{\footnotesize #1}}}}}
\newcommand{\nostep}[2][]{\raisebox{.5pt}{\textcircled{\raisebox{-.5pt}{\small#2\raisebox{0.5pt}{\footnotesize #1}}}}}
\newcommand{\rulesep}{\unskip\ \vrule\ }

\definecolor{attackOne}{HTML}{D7191B}
\definecolor{attackTwo}{HTML}{2B7BB6}
\definecolor{attackThree}{HTML}{FDAE61}
\definecolor{benign}{HTML}{A5A5A5}

\usepackage{subcaption}
\newif\ifmanualheader

\manualheaderfalse

\ifmanualheader
  \makeatletter
  \if@ACM@anonymous
  \else
    \g@addto@macro\@subtitlenotes{\global\let\authors=\cleanauthors}
  \fi
  \makeatother
\fi

\begin{document}

\title[A False Sense of Security? Revisiting the State of Machine Learning-Based Industrial Intrusion Detection]{A False Sense of Security? Revisiting the State of\\Machine Learning-Based Industrial Intrusion Detection}

\ifmanualheader
  \author{Dominik Kus$^{*,\dagger}$, Eric Wagner$^{\dagger,*}$, Jan Pennekamp$^*$, Konrad Wolsing$^{\dagger,*}$,\\ Ina Berenice Fink$^*$, Markus Dahlmanns$^*$, Klaus Wehrle$^{*,\dagger}$, and Martin Henze$^{\ddagger,\dagger}$}
  \def\cleanauthors{Dominik Kus, Eric Wagner, Jan Pennekamp, Konrad Wolsing, Ina Berenice Fink, Markus Dahlmanns, Klaus Wehrle, and Martin Henze}
  \affiliation{
    $^*$\textit{Communication and Distributed Systems}, RWTH Aachen University \country{Germany} $\cdot$ \{lastname\}@comsys.rwth-aachen.de\\ %
    $^\dagger$\textit{Cyber Analysis \& Defense}, Fraunhofer FKIE \country{Germany} $\cdot$ \{firstname.lastname\}@fkie.fraunhofer.de\\
    $^\ddagger$\textit{Security and Privacy in Industrial Cooperation}, RWTH Aachen University \country{Germany} $\cdot$ henze@cs.rwth-aachen.de
  }
\else
  \author{Dominik Kus}
  \email{kus@comsys.rwth-aachen.de}
  \affiliation{%
    \institution{RWTH Aachen University}
    \country{}
  }
 \affiliation{%
	\institution{Fraunhofer FKIE}
	\country{}
}

  \author{Eric Wagner}
  \email{eric.wagner@fkie.fraunhofer.de}
  \affiliation{%
    \institution{Fraunhofer FKIE}
    \country{}
  }
  \affiliation{%
    \institution{RWTH Aachen University}
    \country{}
  }

  \author{Jan Pennekamp}
  \email{jan.pk@comsys.rwth-aachen.de}
  \affiliation{%
    \institution{RWTH Aachen University}
    \country{}
  }

  \author{Konrad Wolsing}
  \email{konrad.wolsing@fkie.fraunhofer.de}
  \affiliation{%
    \institution{Fraunhofer FKIE}
    \country{}
  }
  \affiliation{%
    \institution{RWTH Aachen University}
    \country{}
  }

  \author{Ina Berenice Fink}
  \email{fink@comsys.rwth-aachen.de}
  \affiliation{%
    \institution{RWTH Aachen University}
    \country{}
  }

  \author{Markus Dahlmanns}
  \email{dahlmanns@comsys.rwth-aachen.de}
  \affiliation{%
    \institution{RWTH Aachen University}
    \country{}
  }

  \author{Klaus Wehrle}
  \email{wehrle@comsys.rwth-aachen.de}
  \affiliation{%
    \institution{RWTH Aachen University}
    \country{}
  }
  \affiliation{%
    \institution{Fraunhofer FKIE}
    \country{}
  }

  \author{Martin Henze}
  \email{henze@cs.rwth-aachen.de}
  \affiliation{%
    \institution{RWTH Aachen University}
    \country{}
  }
  \affiliation{%
    \institution{Fraunhofer FKIE}
    \country{}
  }
\fi

\renewcommand{\shortauthors}{Kus et al.}

\begin{abstract}
Anomaly-based intrusion detection promises to detect novel or unknown attacks on industrial control systems by modeling expected system behavior and raising corresponding alarms for any deviations.
As manually creating these behavioral models is tedious and error-prone, research focuses on machine learning to train them automatically, achieving detection rates upwards of~\SI{99}{\percent}.
However, these approaches are typically trained not only on benign traffic but also on attacks and then evaluated against the same type of attack used for training.
Hence, their actual, real-world performance on unknown (not trained on) attacks remains unclear. %
In turn, the reported near-perfect detection rates of machine learning-based intrusion detection might create a false sense of security.
To assess this situation and clarify the \emph{real} potential of machine learning-based industrial intrusion detection, we develop an evaluation methodology and examine multiple approaches from literature for their performance on unknown attacks (excluded from training).
Our results highlight an ineffectiveness in detecting unknown attacks, with detection rates dropping to between \SI{3.2}{\percent} and \SI{14.7}{\percent} for some types of attacks.
Moving forward, we derive recommendations for further research on machine learning-based approaches to ensure clarity on their ability to detect unknown attacks.
\end{abstract}



\maketitle

\section{Introduction}
\label{sec:introduction}

With ongoing digitization, Industrial Control Systems (ICS) are increasingly networked and connected to the Internet~\cite{Pennekampetal2019Towards,Serroretal2021Challenges}, thus suspending the long-deployed air-gap principle as primary protection against intrusion.
However, legacy ICS devices were usually not designed to implement adequate network security and are rarely replaced due to high costs and long device lifetimes~\cite{Krauseetal2021Cybersecurity,Serroretal2021Challenges}.
Consequently, ICS are increasingly targeted by cyberattacks with potentially severe damage~\cite{Huetal2018A,Heetal2016Cyber-physical}, exposing the glaring security deficits of ICS, which stem most importantly from weak or missing protection mechanisms~\cite{McLaughlinetal2016,Abeetal2016Security}.
To alleviate this situation, security mechanisms must be retrofitted for Internet-connected ICS devices.

Network-based intrusion detection systems~(IDSs)~\cite{Open-Information-Security-Foundation-OISF2021Suricata,Sommer2003Bro:} constitute a promising approach for such retrofitting.
While \emph{signature}-based IDSs detect attacks using pre-configured signatures, \eg{} a specific sequence of network packets, \emph{anomaly}-based IDSs model the expected behavior of a system and consider deviations as potential intrusion, \eg{} a control parameter outside physical bounds.
Thus, while signature-based IDSs can only identify known attacks, anomaly-based IDSs promise to also detect novel attacks~\cite{Jyothsnaetal2011A,Tavallaeeetal2010Toward}.

Anomaly-based intrusion detection is particularly well-suited for ICSs, as industrial devices usually exhibit regular and predictable communication patterns~\cite{Hilleretal2018Secure} that remain largely unchanged over time and ease the creation of behavioral models.
However, the individual and application-specific use of industrial devices requires tailoring IDSs to the deployed system, involving high effort~\cite{Almgrenetal2018The}.

A promising approach to address this issue for \emph{industrial} IDSs (IIDSs) is the application of machine learning (ML).
ML algorithms can be trained on historic ICS data, thereby learning properties of the physical system and attacks on it.
Hence, ML algorithms supersede the manual crafting of system models in anomaly detection and signatures in signature-based detection.
Furthermore, the ability of ML to generalize and abstract patterns allows even operating on ``new'' (\ie{} unseen) data.
However, classifying ML-based IDSs as signature- or anomaly-based~\cite{Sommeretal2010Outside}, \ie{} determining whether an IDS learns normal behavior, attack signatures, or both, is non-trivial due to the intransparency of the learning process within ML.

Related work has proposed various ML-based IIDSs, either by training exclusively on benign network traffic from the ICS~\cite{Fengetal2017Multi-level} or by training on a mix of benign and malicious traffic~\cite{Junejoetal2016Behaviour-Based}, indicating almost perfect detection performance in excess of \SI{99}{\percent}~\cite{Lopez-Perezetal2018Machine}.
However, widely-used performance evaluation methods, \eg{} metrics such as precision, recall, or $F_1$-score, only cover the ability of an IIDS to detect known attacks and do \emph{not} capture their ability to detect new variations or even entirely new types of attacks, which is the core promise of anomaly-based over signature-based detection.

We argue that, especially for systems trained on a randomly sampled mix of benign and malicious traffic, it is debatable whether those IIDSs can actually realize anomaly detection as opposed to only learning signatures of trained attacks.
Thus, also taking unseen attacks into account, the \emph{actual} performance of ML-based IIDSs remains unclear to this point.
By only providing evaluation results on known attacks while claiming to perform anomaly detection, such approaches create the impression of almost perfect protection and lead to a false sense of security in real-world deployments.

Thus, clarification of the \emph{real} potential of ML-based IIDSs to offer comprehensive protection is urgently needed.
While literature on attacking specific IIDSs~\cite{Erbaetal2020No} and on performing stealthy attacks against ICS~\cite{Fengetal2017A} exists, there is a lack of research \wrt{} the performance of IIDSs on unknown attacks.
In this paper, we address this issue by examining and evaluating multiple ML-based IIDSs from literature for their performance on unknown attacks.

\noindent\textbf{Contributions.}
Our main contributions are as follows.
\begin{itemize}[noitemsep,topsep=0pt,leftmargin=9pt]
  \item We derive a methodology to evaluate the generalizability of ML-based IIDSs to detect attacks they have not been trained on.
  \item Using our methodology, we evaluate existing ML-based IIDSs on attacks that were deliberately excluded from training.
  Our results show that these approaches only perform well on attacks they have been trained on, leading to a false sense of security.
  \item By further training ML-based IIDSs on only \emph{one} attack type, we assess their ability to work in scenarios where the amount of available training data is limited.
  We show that ML-based IIDSs do not generalize well enough to capture new (unknown) attacks.
\end{itemize}

\noindent\textbf{Availability Statement.}
To foster further research and ensure reproducibility, our evaluation framework and evaluation artifacts are available at: \url{https://github.com/COMSYS/ML-IIDS-generalizability}.

\section{Industrial Intrusion Detection}
\label{sec:background}

Complementing preventive security measures such as firewalls, encryption, and authentication, intrusion detection acts as an important additional safeguard to discover remaining attacks~\cite{Wang2009The,Krauseetal2021Cybersecurity}.
In the following, we provide the necessary background on intrusion detection and argue that its passive nature eases retrofittability and thus makes it especially attractive for industrial networks (Section~\ref{sec:background:ids}).
We then specifically focus on ML-based intrusion detection~(Section~\ref{sec:background:ml}), as it promises strong detection capabilities for advanced attacks on industrial networks~\cite{Bhamareetal2020Cybersecurity,Junejoetal2016Behaviour-Based,Krauseetal2021Cybersecurity}.

\subsection{Traditional vs.\ Industrial IDSs}
\label{sec:background:ids}

Intrusion detection systems (IDSs) passively monitor the behavior of individual devices (host-based) and/or communication between devices (network-based)~\cite{Krauseetal2021Cybersecurity} to discover attacks or suspicious behavior.
To this end, the core idea behind IDSs is that attacks lead to observably different behavior than normal system operation and can thus be detected.
In traditional IDS settings such as office or server networks~\cite{Uetzetal2021Reproducible}, attacks are typically spread out widely, \eg{} through Internet-scale malware.
Rule-based IDSs, such as YARA~\cite{Alvarez2008YARA}, Suricata~\cite{Open-Information-Security-Foundation-OISF2021Suricata}, or Zeek~\cite{The-Zeek-Project2019The} (previously Bro~\cite{Sommer2003Bro:}), can often reliably detect them.
Consequently, state-of-the-art IDSs in these traditional settings often rely on signatures or rules as attack indicators.
They raise an alarm whenever device or communication behavior matches any of the predefined rules or signatures~\cite{Khraisatetal2019Survey}.

Especially in industrial settings, IDSs promise to be an easily deployable safety net for otherwise often insufficiently secured industrial networks~\cite{Serroretal2021Challenges,Dahlmannsetal2020Easing,Dahlmannsetal2022Missed}.
In particular, network-based IDSs can easily be integrated into existing infrastructure without time-consuming and costly changes to deployed devices or software.
Furthermore, the unique characteristics of industrial networks and processes facilitate the use of intrusion detection:
In contrast to traditional IT networks (office or server), communication in industrial networks follows a significantly more regular and predictable pattern~\cite{Hilleretal2018Secure,Wolsingetal2020Poster:}, \eg{} sensor readings that are refreshed with a fixed periodicity~\cite{Akerbergetal2011Future}.
As such, an underlying assumption of industrial IDSs is that attacks likely lead to a distinctly different communication behavior.
At the same time, the strong interdependence between industrial processes and the communication necessary to monitor and control them allows IDSs to detect even subtle attacks, such as minor manipulations to the water's acidity in a water treatment plant~\cite{Urbinaetal2016Limiting}.
Therefore, the industrial context is uniquely suitable for the deployment of, especially anomaly-based, IDSs.

However, while industrial networks provide ample additional opportunities to detect attacks, these attacks typically cannot easily be described using rules or signatures (as it is possible for traditional IT networks)~\cite{Zhouetal2015Design}.
For example, subtle attacks might send unsuspicious network packets (\ie{} those also appearing in legitimate communication), but cleverly time these to bring a supervised process into an unsafe state~\cite{Urbinaetal2016Limiting}.
As detecting such attacks pushes rule- or signature-based IDSs beyond their limits~\cite{Chenetal2018Learning,Junejoetal2016Behaviour-Based}, a large research community has gathered around ML approaches to also detect advanced attacks on industrial networks~\cite{Bhamareetal2020Cybersecurity,Luoetal2022Deep, Wolsingetal2021IPAL:}.

\subsection{Machine Learning for Industrial IDSs}
\label{sec:background:ml}

\begin{figure}
	\centering
	\includegraphics[trim={0pt 56pt 0pt 0pt}, clip]{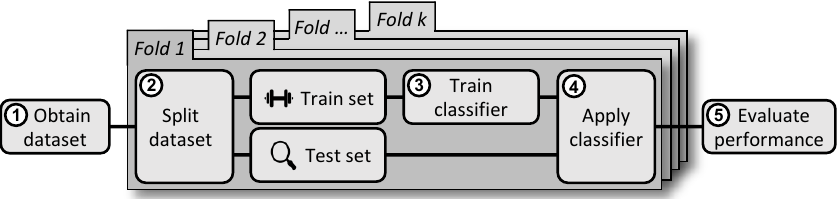}
	\vspace{-2em}
	\caption{%
		$k$-fold cross-validation allows to accurately evaluate the performance of a machine learning classifier.
		For each fold, different parts of the dataset are used for training and testing.
		Eventually, the full dataset is classified (tested) once.
	}
	\Description{%
    The illustrated figure is described in detail in the text.
    It visualizes the order of the reported steps when evaluating the performance of a machine learning classifier.
  }
	\label{fig:background:ml-iids-process}
	\vspace{-1.5em}
\end{figure}

The premise of machine learning (ML) for industrial intrusion detection is simple:
``Automatically'' learn what constitutes benign and malicious behavior to later classify observed behavior without having to care about details~\cite{Sommeretal2010Outside} of the underlying physical process and communication.
We explain the training and evaluation process of an ML-based industrial IDS in the following alongside Figure~\ref{fig:background:ml-iids-process}.
After successful training, the IDS can then be deployed to \emph{predict} whether observed samples (\eg{} a sequence of network packets) constitute an attack.

The first step to creating an ML-based IIDS is to obtain a suitable ICS dataset covering nominal operation \emph{and} labeled attack patterns~(\step{1}).
In \step{2}, this dataset is split into a training (used for creating the classifier) and a testing dataset (used for evaluating the classifier), \eg{} according to a predefined ratio, such as \nicefrac{80}{20}.
The respective training process (\step{3}) depends on the specific classifier, but usually relies on iterative optimization, \eg{} using gradient descent.
After the classifier is trained, in \step{4}, it is applied on the test set to predict whether the contained samples are malicious or not.
Finally, in \step{5}, the prediction results are validated against the known labels (from the dataset) to evaluate the classifier's performance (\ie{} how ``well'' it detects attacks).

In our simplified example, so far, the results are obtained using a single train/test split, \ie testing is performed on a single test set, providing only limited insights into the performance of the IIDS on different datasets.
To mitigate this issue, cross-validation~(CV)~\cite{Refaeilzadehetal2016Cross-Validation} is a structured approach for evaluating an IIDS on multiple train/test splits and thus increases confidence in the results.
For $k$-fold cross-validation, the dataset is partitioned into $k$ equal-sized parts.
As shown in Figure~\ref{fig:background:ml-iids-process}, Steps \nostep{2}--\nostep{4} are then repeated $k$-times using a non-random train/test split where one of the $k$ parts is used as the test set while the remaining $k{-}1$ parts form the train set.
For the final evaluation of the approach, the results from each of those $k$ \emph{folds} are then aggregated, \eg by calculating the arithmetic mean.

While the used classifiers vary widely across approaches (\cf{}~Section~\ref{sec:relatedwork}), the process to measure their performance~(\step{5}) remains the same.
A classifier's predictions over the test set~(\cf{}~\step{4}) are interpreted alongside four possible outcomes:
\emph{True positives~(TP)} count the number of correctly identified malicious samples, while \emph{false positives~(FP)} count the number of benign samples incorrectly classified as malicious.
Analogously, \emph{true negatives~(TN)} count the number of correctly classified benign samples, whereas \emph{false negatives~(FN)} count the number of malicious samples falsely classified as benign.
A perfect classifier has only true positives and true negatives, \ie{} no false positives or false negatives.

Two widely used metrics based on these outcomes are \emph{precision} and \emph{recall}~\cite{Arpetal2022Dos}.
Precision represents the ratio of correctly identified malicious samples among all samples predicted as malicious~(\nicefrac{$\mathit{TP}$}{$(\mathit{TP} + \mathit{FP})$}).
Hence, the optimal value of \num{1} implies that all samples predicted as malicious indeed correspond to malicious behavior (\ie no false alarms).
Recall represents the ratio of correctly identified malicious samples among all actually malicious samples~(\nicefrac{$\mathit{TP}$}{$(\mathit{TP} + \mathit{FN})$}).
For an optimal value of \num{1}, all malicious samples were correctly identified (\ie no malicious sample is missed).
Thus, these metrics complement each other and provide good insights into a classifier's performance, even with unbalanced datasets~\cite{Arpetal2022Dos}.

Optimally, both precision and recall would be \num{1}; however, there is often a trade-off between the two values.
Both metrics can be aggregated into the $F_1$-score (\nicefrac{2}{$(\mathit{Precision}^{-1} + \mathit{Recall}^{-1})$}), which provides a single measure for a classifier's performance.
Ergo, an $F_1$-score of \num{1} would be optimal and imply that both precision and recall are \num{1}.

While employing ML for industrial intrusion detection offers various benefits, such as easier deployment~\cite{Bhamareetal2020Cybersecurity}, one fundamental drawback is that classifiers can only ``learn'' information and patterns for which corresponding training data exists~\cite{Sommeretal2010Outside}.
Consequently, while ML promises to also detect advanced attacks on industrial networks that rules or signatures cannot cover, there is a latent apprehension that such approaches can only detect attacks they have been trained on and thus remain oblivious of all other, especially evolving and newly developed, attacks.

\textbf{Takeaways.}
Intrusion detection, especially based on ML, is promising to easily retrofit industrial networks with capabilities to timely uncover attacks.
However, there is an inherent risk that ML-based IDSs for industrial networks cannot generalize to detect attacks they have not been trained on.

\begin{figure*}[t]
  \centering
  \begin{subfigure}[c]{.32\textwidth}
  \centering
  \includegraphics[trim={0pt 6pt 0pt 0pt}, clip]{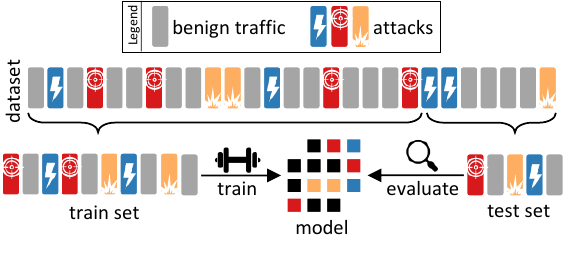}
  \vspace{-1.2em}
  \caption{%
    Traditional ML approach, using a \textcolor{attackOne}{ra}\textcolor{benign}{nd}\textcolor{attackTwo}{om} \textcolor{benign}{sa}\textcolor{attackThree}{mp}\textcolor{benign}{le} of the dataset for classifier training.
  }
  \label{fig:design:traditional}
  \end{subfigure}
  \rulesep
  \begin{subfigure}[c]{.32\textwidth}
  \includegraphics[trim={0pt 6pt 0pt 0pt}, clip]{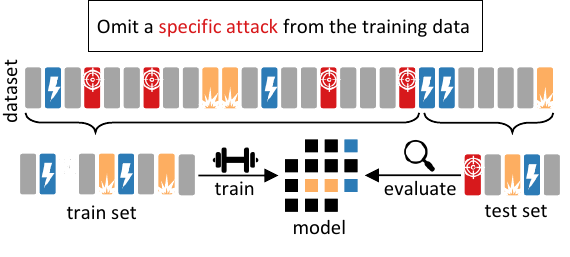}
  \vspace{-1.2em}
  \caption{%
    Excluding a \textcolor{attackOne}{specific attack} during the training of the classifier to test its generalizability.
  }
  \label{fig:design:omit}
  \end{subfigure}
  \rulesep
  \begin{subfigure}[c]{.32\textwidth}
  \includegraphics[trim={0pt 6pt 0pt 0pt}, clip]{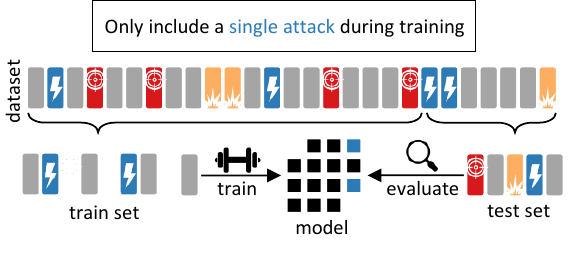}
  \vspace{-1.2em}
  \caption{%
    Focusing on a \textcolor{attackTwo}{single attack} only to reveal interrelations between different trained attacks.
  }
  \label{fig:design:single}
  \end{subfigure}
  \vspace{-0.5em}
  \caption{%
    Traditional evaluations source their training data from a random sample.
    Thus, they fail to properly test the classifier's generalizability.
    By proposing a methodology consisting of two experiments, we intend to address this gap in IIDS research.
  }
  \Description{%
    The figure details three ways how a dataset can be split into train and test sets.
    The utilized dataset contains benign data and three types of attacks, each with multiple repetitions.
    In approach (a), which is the traditional ML approach, all three types of attacks end up in both train and test set.
    In approach (b), a specific attack is omitted from training, such that the test set contains all attack types while the train set is missing one of them.
    In approach (c), the classifier is trained while having access to a single attack only, such that the test set contains all attack types while the train set only contains one type of attack.
  }
  \label{fig:design:methodology-overview}
  \vspace{-0.5em}
\end{figure*}

\section{Related Work on ML-Based IIDSs}
\label{sec:relatedwork}

In recent years, ML, with all its benefits and potential pitfalls, has seen widespread application in industrial intrusion and anomaly detection, and a plethora of ML-based anomaly detectors have been proposed~\cite{Bhamareetal2020Cybersecurity,Luoetal2022Deep,Umeretal2022Machine}.
Notably, supervised learning-based approaches gained significant interest for their high attack and anomaly detection rates~\cite{Ramanetal2021Machine}.
Such supervised classifiers include random forests (RFs)~\cite{Al-Abassietal2020An,Junejoetal2016Behaviour-Based,Lopez-Perezetal2018Machine,Kumaretal2021Machine,Colellietal2021Anomaly-Based,Antonetal2019Anomaly-based,Anthietal2021A,Aroraetal2021Evaluation,Mokhtarietal2021A}, which utilize multiple decision trees to split a dataset's features into similar classes.
Contrary, support vector machines (SVMs)~\cite{Junejoetal2016Behaviour-Based,Kharitonovetal2019Intrusion,Lopez-Perezetal2018Machine,Kumaretal2021Machine,Antonetal2019Anomaly-based,Chenetal2018Learning,Anthietal2021A,Aroraetal2021Evaluation} map all features into a vector space and derive decision boundaries to separate individual classes.
From a different angle, neural networks~(NNs)~\cite{Al-Abassietal2020An,Chuetal2019Industrial,Junejoetal2016Behaviour-Based,Kharitonovetal2019Intrusion,Lopez-Perezetal2018Machine,Radoglou-Grammatikisetal2020DIDEROT:,Homayounietal2020An,Fengetal2017Multi-level,Ramanetal2019Anomaly,Gohetal2017Anomaly} mimic human brains and can be trained to model any function, e.g., classifying input features as benign or malicious.
Besides this plurality in ML-based approaches, their common motivation is to increase utility compared to deterministic signature-based intrusion detection through
\begin{enumerate*}[label=(\roman*)]
	\item generalizability across domains and
	\item the ability to identify novel, not previously seen, anomalies.
\end{enumerate*}

However, whether machine-learning intrusion detection can indeed live up to these promises has been challenged recently~\cite{Erbaetal2020No,Wolsingetal2021IPAL:, Ahmedetal2020Challenges,Ramanetal2021Machine}.
Concerning generalizability, recent investigations~\cite{Erbaetal2020No,Wolsingetal2021IPAL:} show that applying IIDSs to novel domains is not as straightforward, and its success depends on the underlying detection methods.
With respect to the ability to detect unknown, \ie not trained on, attacks, even if detection scores for individual attack types are calculated~\cite{Anthietal2021A,Chenetal2018Learning,Chuetal2019Industrial,Fengetal2017Multi-level,Junejoetal2016Behaviour-Based,Ramanetal2019Anomaly,Lopez-Perezetal2018Machine}, it remains unclear how classifiers handle unseen attacks or variations.
The claimed strength of many ML-based IIDSs can be further questioned when considering that classifiers that only train on benign data~\cite{Aoudietal2018Truth,Linetal2018TABOR:,Fengetal2019A} generally suffer from worse detection performance and higher false-positive rates~\cite{Junejoetal2016Behaviour-Based}.
While these discussions point out problems \wrt{} claimed generalizability of existing results beyond specific evaluation scenarios, they do not further focus on the underlying evaluation methodology in initially proposed scenarios.

To this end, problems with the evaluation methodology of ML-based systems in general and anomaly detectors in specific were raised by various related work~\cite{Sommeretal2010Outside,Arpetal2022Dos,Ahmedetal2020Challenges,Ramanetal2021Machine,Tavallaeeetal2010Toward,Pendleburyetal2019TESSERACT:,Umeretal2022Machine}.
For one, the interpretation of evaluation results is not straightforward, leading to false conclusions from misinterpreted data~\cite{Arpetal2022Dos,Pendleburyetal2019TESSERACT:}.
Moreover, related work questions whether ML is actually suitable for anomaly detection, or if it is rather only able to detect variations of already seen attacks in office networks~\cite{Sommeretal2010Outside} as well as for ICSs~\cite{Ahmedetal2020Challenges,Ramanetal2021Machine}.
However, these statements are not backed with detailed analysis of the actual ability of ML to perform anomaly detection for ICS networks, where the predictable behavior under normal operation potentially allows to still detect anomalies.
In a similar vein, the enhancement of existing datasets with additional, artificial attack signatures to more accurately present what constitutes anomalous behavior is proposed~\cite{Wijayaetal2020Domain-Based, Umeretal2021Attack}.
However, the exact implications for ML-based IIDSs that train on those enhanced datasets remain unexplored.

\textbf{Takeaways.}
Machine learning-based industrial intrusion detection received broad attention from research due to the promise to generalize across domains and detect previously unseen anomalies.
Simultaneously, it becomes clear that such claims require much more scrutiny.
While first detailed analyses of the generalizability of industrial intrusion detection have been conducted~\cite{Erbaetal2020No,Wolsingetal2021IPAL:}, to the best of our knowledge, the potential of (supervised) ML to detect unknown, \ie not trained on, anomalies remains unexplored.
Thus, it remains unclear whether novel anomalies and variations of known attacks can be detected reliably by ML, demanding a methodology to perform such an analysis.

\section{Dissecting ML-based IIDS Evaluations}
\label{sec:design}

The main goal of evaluating an IDS is to gain insight into its capabilities~\cite{Sommeretal2010Outside}.
While performance metrics help to express achieved results, the underlying evaluation methodology is far more important to accurately rate IDS performance, as it dictates how to interpret generated metrics and what meaning they convey for practical, real-world performance.
Given safety-related considerations in industrial settings, this issue is crucial for deployed IIDSs.

Due to the specific abstracting properties of ML, the underlying methodology is especially relevant for ML-based IIDSs.
When surveying ML-based security evaluations, research discovered various prevalent pitfalls in a multitude of papers~\cite{Arpetal2022Dos}.
In our context, the selection of the test dataset is most concerning as it determines what scenarios the system is actually tested for and what scenarios are not credibly covered by the proposed IIDS.

To shine a light on this issue, we discuss the traditional, state-of-the-art evaluation methodology and its deficits \wrt{} the IIDSs' ability to generalize to new attacks (Section~\ref{sec:design:todaysoverview}).
We address those shortcomings by proposing an improved evaluation methodology that is specifically tailored to evaluating ML-based IIDSs (Section~\ref{sec:design:experiments}).

\subsection{Today's Traditional Evaluation Approach}
\label{sec:design:todaysoverview}

A commonly used approach to evaluate ML-based IIDSs is to apply a traditional evaluation methodology from the field of ML, as we illustrate in Figure~\ref{fig:design:traditional}.
Here, the final evaluation is performed on the test set, which corresponds to a sample of the original dataset that was withheld during training.
Thus, the classifier has not seen this exact data during the training step.
Based on the classification results of the test set, metrics, such as precision and recall, express the achieved performance.
We refer to Section~\ref{sec:background:ml} for more elaborate details on the general training and evaluation processes.

While precision- and recall-based metrics are generally recommended to address an imbalanced dataset and to prevent the base rate fallacy~\cite{Arpetal2022Dos}, the selection of the test set warrants further attention.
When randomly sampling both training and testing data from the ICS source dataset, they most likely cover similar observations.
Hence, they are very homogeneous (visualized by the same types of attacks in both train and test set in Figure~\ref{fig:design:traditional}).
Thus, assuming the measured performance to be indicative of real-world performance contains the implicit assumption that the real-world environment in which the IIDSs will be deployed is homogeneous to the train set---an assumption with severe implications.

While this assumption is reasonable for traditional ML applications, such as image classification or natural language processing, it is not suitable for security-related applications, and more specifically IIDSs, for the following reasons:
\begin{enumerate*}[label=(\roman*)]
  \item Identifying and predicting all perceivable risks is highly unlikely, and
  \item attackers are constantly adapting and improving their tools.
\end{enumerate*}
Consequently, IIDSs are most likely confronted with novel attacks or variants of existing attacks during their deployment.
Regardless, those scenarios are not captured by the above methodology:
Given that the sampled test and train sets are homogeneous, the likelihood of new, unknown attacks is not accounted for while testing the classifier.

Unfortunately, for this reason, corresponding evaluations leave many questions unanswered.
In particular, they fail to explain how systems react to new attacks and what their overall classification limitations are.
However, these aspects are important to provide accurate evaluations~\cite{Sommeretal2010Outside}.
Thus, today's inadequate evaluation methodology can lead to a warped perception of IIDS performance and create a false sense of security.
To tackle this problem and to develop further insight into the IIDSs' capabilities, especially concerning their ability to detect unknown attacks in the wild, we propose an improved evaluation methodology for ML-based IIDSs.

\subsection{Our Proposed Evaluation Methodology}
\label{sec:design:experiments}

In the following, we introduce our new methodology to allow for expressive performance evaluations in the context of ML-based IIDSs and thus address today's methodical shortcomings.

\subsubsection{Methodology Overview}
\label{sec:design:methodologyoverview}

Overall, we propose a methodology to evaluate the performance of ML-based IIDSs that sources from two distinct experiments.
Thereby, we intend to provide additional insights into the real-world performance of deployed IIDSs.
In the first experiment, we investigate the IIDSs' ability to detect new attacks by deliberately omitting attacks from the training set.
This setting synthetically simulates a situation where the IIDS is confronted with a new, unknown kind of attack during its deployment.
For the second experiment, we only train the classifier for a single kind of attack (as well as benign data) by omitting other (known) attacks from the training set.
This setting can help to put the results from the first experiment into perspective.
It further highlights interrelations between different attacks present in the dataset.
Finally, for settings where the intended generalization failed to reliably detect (new) attacks, it answers the question of whether ``specialized'' classifiers that only focus on single attacks are a promising way.

Our methodology is independent of the used classifier (\ie used IIDS) and the input dataset (which only must be segmentable into different classes of attacks).
In this paper, we systematically analyze and compare the performance of three different ML-based IIDS classifiers (\cf Section~\ref{sec:evaluation}).
To ensure generalizable and significant results, we also mandate the use of cross-validation for our methodology.
In Figures~\ref{fig:design:omit} and~\ref{fig:design:single}, we visualize how the train and test sets are prepared.
Now, we elaborate on the individual experiments.

\subsubsection{Detection of New, Unknown Attacks}
\label{sec:design:unknownattacks}

Overall, we propose a method to analyze a classifier's ability to detect novel attacks during its deployment.
To this end, we specifically adapt the dataset splitting (\step{2} in Figure~\ref{fig:background:ml-iids-process}) prior to the classifier's training (\step{3}).
We illustrate the general idea for a single fold in Figure~\ref{fig:design:omit}.
The subsequent performance evaluation of the used classifier is then based on known metrics, \ie precision and recall.

As a prerequisite, the dataset is split into $k$ parts of equal size, which corresponds to a $k$-fold cross-validation, \ie we use a single part for the test set while the remaining parts constitute the train set.
When using $k=5$, the resulting train set corresponds to approximately \SI{80}{\percent} of the input data, and the remaining \SI{20}{\percent} are used for the test set.
Subsequently, for each type of attack, we conduct an evaluation.
More specifically, we consider every attack to be unknown once, \ie we filter all instances of this attack from the train set and move them to the test set.
While this approach alters the ratio between train and test set, its implications are usually negligible in practice due to the inherent imbalance between benign and malicious packets (in common ICS datasets).
Thus, we prepare $k$ folds for each type of attack where the instances related to the attack are only part of the test set, \ie{} overall, we repeat the process of Figure~\ref{fig:design:omit} $(k \times \text{\#attack})$ times to evaluate a classifier.

After training and testing on all these folds, we individually compute the performance metrics (typically, precision and recall) for each type of attack.
Inspecting those results can provide an overall impression of how well the classifier abstracts from the specifics of the dataset, \eg by learning process-specific parameters or communication patterns of the ICS, to detect attacks without relying on specific pre-trained attack patterns.
Thus, our methodology provides an understanding of whether the ML-based IIDS \emph{only} detects known attacks (\ie being limited in its practical use), or whether it is also able to generalize the input data---a central promise of ML-based approaches---\ie to also reliably detect ``anomalies'' in safety-critical, real-world deployments.

\subsubsection{Independently Evaluating Attacks}
\label{sec:design:singleattacks}

Following this first experiment, we intend to obtain an improved understanding of how the classifier learns, deals with, and abstracts from seen (trained on) attacks.
For evaluations, these insights are instrumental in different ways:
\begin{enumerate*}[label=(\roman*)]
  \item They underline why we observe a generalizability for specific types of attacks,
  \item they allow us to identify interrelations between attacks, and
  \item thus, they also assist in understanding results that we obtained by performing the first experiment.
\end{enumerate*}

We propose a second experiment that focuses only on individual attacks.
To this end, we again adapt the dataset splitting (\step{2} in Figure~\ref{fig:background:ml-iids-process}).
When compared to the first method, the overall setup is very similar:
We only modify the applied filter as we detail in Figure~\ref{fig:design:single} to only keep one attack at a time in the train set.
In particular, for each attack, we fully exclude all other attacks from the train set, \ie{} we filter all malicious instances not belonging to the respective attack from the train set and move them to the test set.
Thus, we train a classifier on a single type of attack.
For a complete evaluation, we repeat this process $(k \times \text{\#attack})$ times.

Overall, the first experiment corresponds to an ``all-but-one'' approach and the second experiment follows an ``only-one'' evaluation.

While we primarily want to study the classifier concerning specific attacks, the observed results might also reveal interrelations between different types of attacks in the dataset.
In addition, this experiment can highlight issues of the classifier related to underfitting, \ie{} the used classifier is not able to properly handle all types of attacks in the dataset simultaneously.
Thus, it provides important insights into today's challenges with ML-based IIDSs.
To address the issue of underfitting, a potential approach could be to also train a ``specialized'' classifier for this specific, challenging attack only.

\textbf{Takeaways.}
Today's state-of-the-art evaluation methodology for ML-based IIDSs fails to consider the approaches' ability to detect new, unknown attacks.
As this property is crucial to properly assess their capabilities (especially in light of generalization), we propose a methodology consisting of two experiments.
Thereby, we are able to obtain further insights into how IIDSs deal with attacks and how well they generalize to novel forms of attacks, \ie how much security they can provide in real-world settings.

\section{Revisiting the Evaluation of IIDSs}
\label{sec:evaluation}

To assess the impact of shortcomings in today's evaluation methodology for ML-based IIDSs (\cf{} Section~\ref{sec:design:todaysoverview}) and shine a light on their true capability to detect novel attacks, we apply our methodology proposed in Section~\ref{sec:design:experiments} to three ML-based IIDSs from literature.

To this end, we first discuss our experimental setup, including the examined IIDSs, the used dataset, and the specific application of our methodology (Section~\ref{sec:evaluation:setup}).
The results of omitting certain attacks or attack categories from training show that the examined approaches are largely unable to detect novel attacks (Section~\ref{sec:evaluation:unknownattacks}).
Training the IIDSs on single attacks or categories reveals that cross-learning between attacks and categories is limited to some special cases.
Classifiers mostly learn the attacks on which they are explicitly trained on.
Combining the results from both experiments details that the examined IIDSs, despite contrary promises, behave much more like signature-based than anomaly-based IDSs (Section~\ref{sec:evaluation:singleattacks}).

\subsection{Experimental Setup}
\label{sec:evaluation:setup}

Over the last years, a plethora of ML-based IIDSs have been proposed and evaluated without much concern about how these generalize to new attacks~(\cf~Section~\ref{sec:relatedwork}).
In this paper, we specifically revisit the performance of three recently proposed IIDSs based on different ML algorithms: RFs, SVMs, and BLSTMs~\cite{Lopez-Perezetal2018Machine}.
These IIDSs constitute prime candidates for our analysis as
\begin{enumerate*}[label=(\roman*)]
  \item they feature official open-source implementations, and
  \item research has independently validated and reproduced their reported performance~\cite{Wolsingetal2021IPAL:}.
\end{enumerate*}
Specifically, for our evaluation, we rely on existing IIDS implementations that utilize the industrial abstraction layer IPAL~\cite{Wolsingetal2021IPAL:}.

While our experiments arguably would benefit from a larger set of IIDSs, we were unable to consider additional implementations.
Due to the lack of publicly available artifacts, we initially tried to contact the authors of four recent publications.
Unfortunately, we only received a single negative response, indicating that the main author no longer works at the corresponding lab.
Subsequently, we attempted to re-implement these approaches on our own.
However, despite this effort, we were unable to reproduce the reported results.
Regardless, given the variety of our considered ML algorithms, covering both traditional ML~(RFs and SVMs) as well as deep learning techniques~(BLSTMs), we are confident to report representative results in this paper.
In the future, other researchers can easily repeat our experiments for other classifiers due to the use of IPAL.

\begin{table}
  \centering
  \caption{Attack categories as introduced for the gas pipeline dataset~\cite{Morrisetal2015Industrial} that we use during our methodology evaluation.}
  \label{tab:evaluation:dataset-categories}
  \vspace{-0.5em}
  \fontsize{8.3}{11}\selectfont
  \begin{tabular}{cllc}
    \toprule
    \textbf{ID} & \textbf{Abbr.} & \multicolumn{1}{c}{\textbf{Descriptive Name}} & \textbf{\#Attacks} \\
    \midrule
    1 & NMRI & Na\"ive Malicious Response Injection& 4 \\
    2 & CMRI & Complex Malicious Response Injection & 7 \\
    3 & MSCI & Malicious State Command Injection & 5 \\
    4 & MPCI & Malicious Parameter Command Injection & 12 \\
    5 & MFCI & Malicious Function Code Injection & 3 \\
    6 & DoS & Denial of Service& 1\\
    7 & Recon & Reconnaissance & 3 \\
    \bottomrule
  \end{tabular}
  \vspace{-2.0em}
\end{table}

For our evaluation, we require datasets that ideally contain multiple instances of labeled attack types (\cf Section~\ref{sec:design:experiments}).
In particular, in this paper, we rely on an established dataset from a gas pipeline ICS that has been specifically designed for cybersecurity research~\cite{Morrisetal2015Industrial}.
At the hardware level, the ICS consists of a pressure sensor and two actuators (a pump and a solenoid valve) that are automated by a control system to regulate the pipeline's pressure.
In total, the dataset contains \num{274628} network packets, out of which about \SI{22}{\percent} belong to labeled attacks.
These attacks stem from \num{35} different attack types designed for this ICS, which are grouped into \num{7}~categories, as we detail in Table~\ref{tab:evaluation:dataset-categories}.

For \emph{NMRI}- and \emph{CMRI}-related attacks, the attacker injects malicious sensor readings and setpoints to manipulate control algorithms.
Attacks that are included in \emph{MSCI}, \emph{MPCI}, and \emph{MFCI} send manipulated commands to actuators to change the state of devices or interfere with their communication.
The \emph{DoS} category consists of a single type of attack in which bad CRC checksums are used to disrupt a device's functionality by overloading it with invalid messages.
Finally, the \emph{Recon}-related attacks attempt to obtain insights into the ICS's operation by scanning for devices and their functionality.
For the ordering of attacks within a category, we follow the original publication.

For both described experiments (\cf Section~\ref{sec:design:experiments}), we conducted our evaluation on two different aggregation levels:
\begin{enumerate*}[label=(\alph*)]
  \item by omitting/training all attacks from a specific attack category (\cf Table~\ref{tab:evaluation:dataset-categories}), and
  \item by focusing our analysis on the individual \num{35} attack types to analyze the generalization \emph{within} and \emph{between} attack categories.
\end{enumerate*}
In line with best practices, we use \num{5}-fold cross-validation throughout our evaluation to retain the \nicefrac{80}{20}-split between train and test set from the original publication~\cite{Morrisetal2015Industrial}.
We report on the average recall and precision values across our \num{5}~folds and forgo examining the differences between the folds as they are irrelevant for our analysis.

\subsection{Understanding the Impact of Novel Attacks}
\label{sec:evaluation:unknownattacks}

\begin{figure}[t]
  \centering
  \includegraphics[width=\columnwidth]{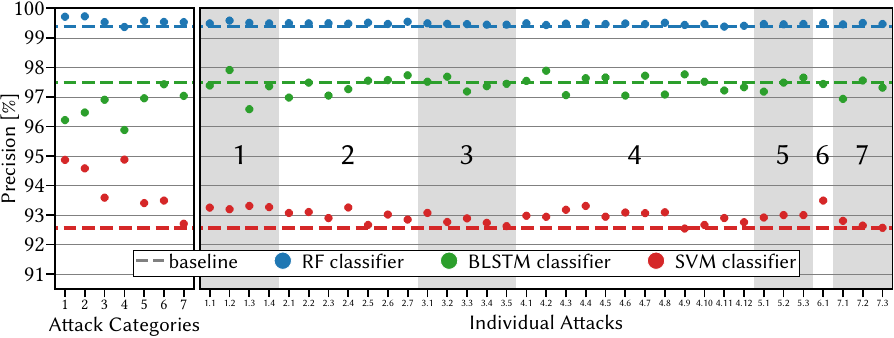}
  \vspace{-2em}
  \caption{%
  Precision over the test set changes only minimally when attack categories (left and shaded in on the right) or individual attacks (right) are omitted from training.
  Notably, BLSTM's precision falls below the baseline in many cases, while the precision for RF and SVM generally exceeds it.
  }
  \Description{%
    A scatter plot that shows the precision between 90 and 100 \% on the Y axis and the removed individual attacks as well as categories on the X axis.
    Three series are shown for the three evaluated classifiers RF, BLSTM and SVM as well as a baseline for each of them.
    The baselines are at around 99.5 \% (RF), 97.5 \% (BLSTM) and 92.5 \% (RF) respectively.
    RF's precision stays close to its baseline but slightly increases in most cases.
    BLSTM's precision is more scattered, decreasing by up to 2 \% for omitted categories while fluctuating up to 1 \% above or below the baseline for omitted attacks.
    SVM's precision increases up to 2 \% over its baseline with most increases being around 0.5 \%.
  }
  \label{fig:evaluation:precision}
  \vspace{-1.5em}
\end{figure}

\begin{figure*}[t]
  \centering
  \subcaptionbox{RF classifier\label{fig:evaluation:unknown:categories-rf}}[.32\textwidth][c]{%
    \includegraphics[width=.32\textwidth]{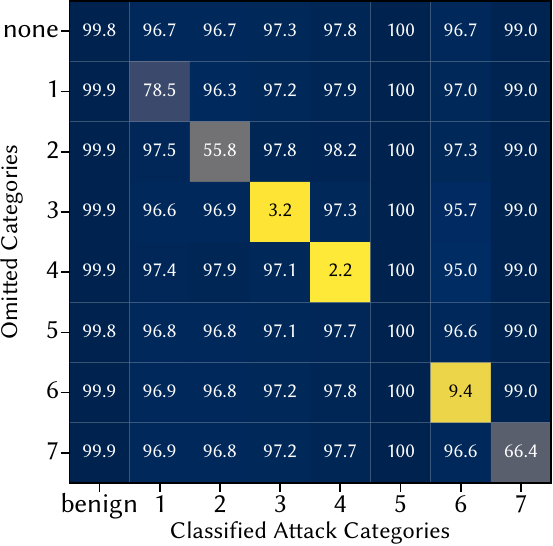}}\quad
  \subcaptionbox{BLSTM classifier\label{fig:evaluation:unknown:categories-blstm}}[.32\textwidth][c]{%
    \includegraphics[width=.32\textwidth]{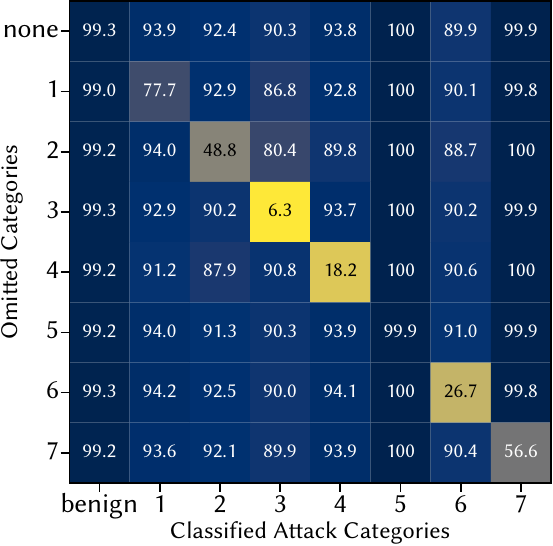}}\quad
  \subcaptionbox{SVM classifier\label{fig:evaluation:unknown:categories-svm}}[.32\textwidth][c]{%
    \includegraphics[width=.32\textwidth]{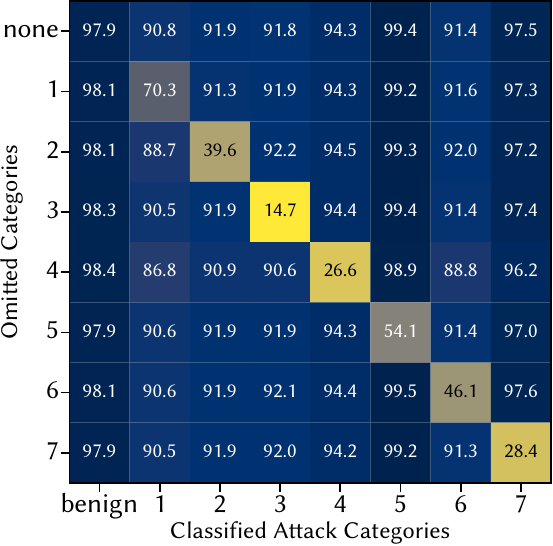}}

  \includegraphics[width=.9\textwidth]{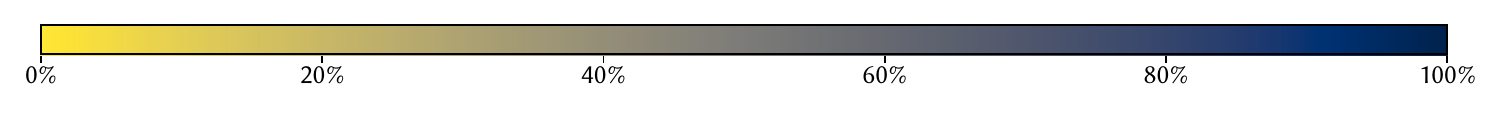}
  \vspace{-0.5em}

  \subcaptionbox{RF classifier\label{fig:evaluation:unknown:attacks-rf}}[.32\textwidth][c]{%
    \includegraphics[width=.32\textwidth]{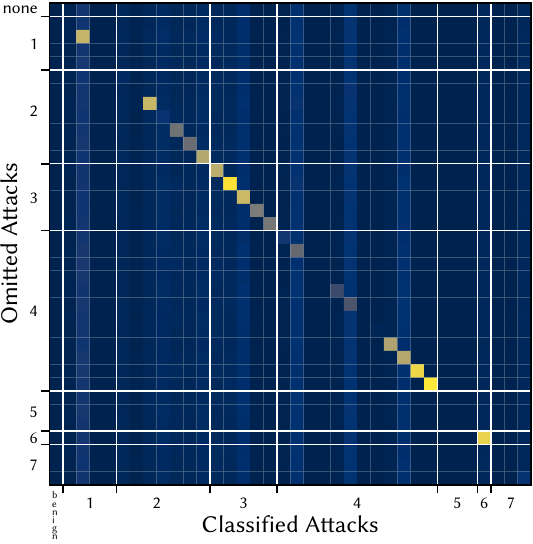}}\quad
  \subcaptionbox{BLSTM classifier\label{fig:evaluation:unknown:attacks-blstm}}[.32\textwidth][c]{%
    \includegraphics[width=.32\textwidth]{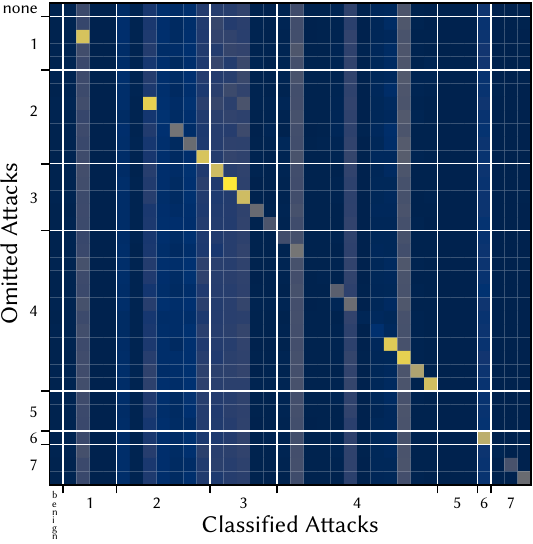}}\quad
  \subcaptionbox{SVM classifier\label{fig:evaluation:unknown:attacks-svm}}[.32\textwidth][c]{%
    \includegraphics[width=.32\textwidth]{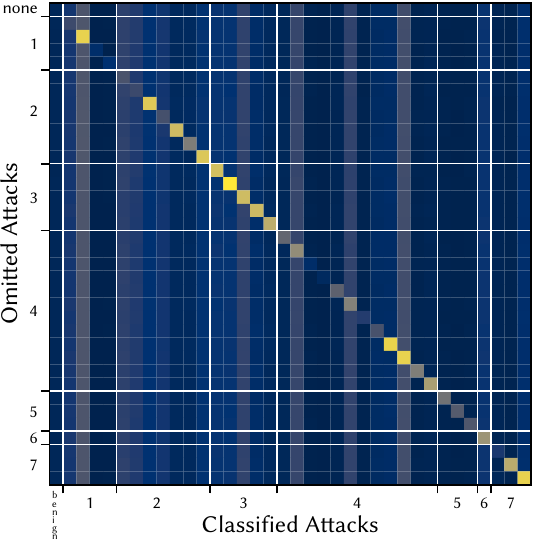}}
  \caption{%
    We applied our evaluation methodology (Section~\ref{sec:design:unknownattacks}) to RF, BLSTM and SVM classifiers from literature on two aggregation levels: once by omitting attack categories (\subref{fig:evaluation:unknown:categories-rf}--\subref{fig:evaluation:unknown:categories-svm}) and once by omitting individual attacks (\subref{fig:evaluation:unknown:attacks-rf}--\subref{fig:evaluation:unknown:attacks-svm}).
    The achieved recall [\%] is visualized as a heatmap where each row corresponds to one omitted attack or category, and each column matches the attack or category for which the recall is measured, respectively.
    The results illustrate a significant drop in recall for many cases, indicating that those attacks or categories are hardly detected by the used IIDS when not explicitly trained on.
  }
  \Description{%
    Deviations from the baseline are almost exclusively observed on the main diagonal of the heatmaps, meaning within the category that was omitted.
    Most of those fields indicate a big drop in recall while some stay close to the baseline.
    A detailed description and analysis is available in the text.
  }
  \label{fig:evaluation:unknown}
  \vspace{-.8em}
\end{figure*}

In our first analysis~(\cf~Section~\ref{sec:design:unknownattacks}), we investigate how the three ML-based IIDSs perform when they are challenged to classify novel, previously unseen, attacks.
To this end, we omit individual attacks or entire categories one by one during training.
As established earlier~(\cf~Section~\ref{sec:background:ml}), we have to consider both precision and recall to gain a complete understanding of a classifier's performance.

In Figure~\ref{fig:evaluation:precision}, we thus analyze the effects on precision first.
For the RF- and SVM-based classifiers, we observe a slight improvement, indicating that the classifiers can separate benign and malicious traffic easier if the trained on attacks are less diverse.
Contrary, the BLSTM performance decreases by up to \num{1.6} percentage points if certain attacks are not seen during training, especially if entire attack categories are not trained on, indicating difficulties identifying benign behavior.
In general, the effects on precision when omitting certain attacks from the train sets are, however, rather marginal.

Besides precision as a performance metric, recall is the key metric for our evaluation (and real-world deployments) as it allows measuring how the detection of an attack type or category changes when it is omitted from training.
Optimally, we would expect an anomaly detector to retain good recall for an attack even if it is omitted from training, indicating its ability to generalize beyond the known attacks.
Next, we discuss how well ML-based IIDSs actually detect novel attacks, as well as variations of known attacks.

\subsubsection{Omitting Entire Attack Categories}

First, we analyze the recall when entire attack categories are omitted, which we illustrate as the recall value heatmaps for the three classifiers in Figures~\ref{fig:evaluation:unknown:categories-rf}--\ref{fig:evaluation:unknown:categories-svm}.
We further include the baseline of training on all attack categories as well as the changed recall when removing a single attack category during training.
Each row corresponds to an experiment where the category, denoted on the y-axis by its ID (\cf Table~\ref{tab:evaluation:dataset-categories}), was omitted.
Additionally, the first row~(``none'') denotes the baseline where the whole train set was used.
Each column again corresponds to an attack category, but now with the first column~(``benign'') denoting benign traffic.
Each field of the heatmap shows the recall averaged over $5$~folds.
The value of \SI{93.7}{\percent} in the fourth row and the fifth column of, \eg Figure~\ref{fig:evaluation:unknown:categories-blstm} thus indicates that \SI{93.7}{\percent} of network packets in the test set belonging to Attack Category~4 were detected by the BLSTM classifier when it was not trained on Category~3.

Throughout Figures~\ref{fig:evaluation:unknown:categories-rf}--\ref{fig:evaluation:unknown:categories-svm}, we observe drops in the recall values primarily on the main diagonals of the heatmaps falling from, \eg{} \SI{90.3}{\percent} to just \SI{6.3}{\percent} in Category~3 using the BLSTM classifier.
These results are expected as those fields correspond to the recall of attacks that were omitted from training.
However, the severity varies widely between categories and classifiers:
Categories~3,~4, and~6 are most severely affected, with drops in recall between \num{45} and \num{94} percentage points, reaching a recall of as low as \SI{2.2}{\percent} compared to the baseline of \SI{97.8}{\percent} for Category~4 and the RF classifier.
Except for the SVMs, all values in those categories fall below \SI{30}{\percent} from a baseline of over \SI{90}{\percent}.
Categories~1,~2, and~7 undergo a smaller drop between \num{16} and \num{52} percentage points, reaching as low as \SI{39.6}{\percent} for Category~2 in the SVM classifier.
For the SVM classifier, Category~7 is an exception as its recall drops by almost \num{70} percentage points to just \SI{28.4}{\percent}.
Finally, Category~5 is an outlier as RF and BLSTM still recognize it with a recall of almost \SI{100}{\percent}, even when not being trained on, showing virtually no performance degradation compared to the baseline.
With the SVM, however, the performance decreases more significantly, achieving a recall of only \SI{54.1}{\percent}.

The primary, but incomplete, explanation for the recall reduction is that
\begin{enumerate*}[label=(\roman*)]
  \item the classifiers have mostly learned signatures of attacks in contrast to the repetitive normal behavior of the ICS, and
  \item it depends on the amount of overlap between the omitted and other remaining categories.
\end{enumerate*}
As such, Categories~1 and~2, both representing different forms of malicious response injection (\cf{} Section~\ref{sec:evaluation:setup}), contain multiple attacks that manipulate the reported pressure value in different ways, which is a plausible explanation for the classifiers to correlate both categories.
However, this explanation would imply similar behavior across different classifiers.

While the observed drop in recall across the classifiers is similar in magnitude for most categories, there are some notable differences.
In particular, a classifier's ability to retain a decent recall for Categories~1,~2,~5, and~7 seems to correlate with an inability to retain it for Categories~3,~4, and~6.
While the SVM classifier fares worse on Categories~1,~2,~5, and~7, being consistently outperformed with at least \num{9} percentage points by the RF and BLSTM classifiers, it outperforms them in Categories~3,~4, and~6, where it reaches a higher recall by between \num{8} and \num{20} percentage points.
Similarly, the BLSTM classifier is outperformed by RF on Categories~1,~2,~5, and~7 while outperforming it on Categories~3,~4, and~6.
It seems that classifiers, which model known attacks accurately and thus achieve high recall on them (\eg RF-based classifiers), tend to specifically overfit those known attacks and are therefore less likely to detect different kinds of anomalies.
This observation indicates the existence of a major difference between the classifiers' general or specific understandings of anomalous behavior.
Most indicative of this phenomenon is the detection of attacks from Category~5 when it is not learned:
While the RF- and BLSTM-based classifiers are perfectly capable of identifying this traffic as malicious, with recall values of \SI{100}{\percent} and \SI{99.9}{\percent}, respectively, the SVM classifier is only able to label about half (\SI{54.1}{\percent}) of the corresponding malicious traffic correctly.

Finally, we examine the recall observed aside from the main diagonal, which mostly details little variation.
However, cases exist where the recall of a specific attack category increases as a different category is omitted, \eg{} the recall for Category~6 \textit{increases} from \SI{96.7}{\percent} to \SI{97.3}{\percent} when Category~2 is omitted for the RF classifier.
Contrary, there are cases where we notice the opposite observation, \eg{} as the recall for Category~6 \textit{decreases} from \SI{96.7}{\percent} to \SI{95.0}{\percent} when Category~4 is omitted when training the same classifier.
The size of those fluctuations seems to be dependent on the classifier:
While the recall in Category~3 changes by at most \num{1.2} percentage points when omitting a different category during the training of the RF and SVM classifiers, it drops by nearly \num{10} percentage points when Category~2 is omitted for the BLSTM classifier.
Overall, we can thus conclude that not training for a specific category in general only influences the detection rate of that specific category, with a mostly insignificant influence on the detection rate of related categories.

\begin{figure*}[t]
  \centering
  \subcaptionbox{RF classifier\label{fig:evaluation:single:categories-rf}}[.32\textwidth][c]{%
    \includegraphics[width=.32\textwidth]{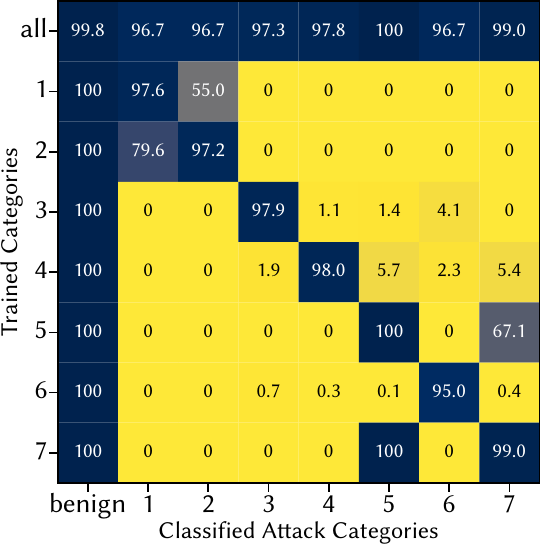}}\quad
  \subcaptionbox{BLSTM classifier\label{fig:evaluation:single:categories-blstm}}[.32\textwidth][c]{%
    \includegraphics[width=.32\textwidth]{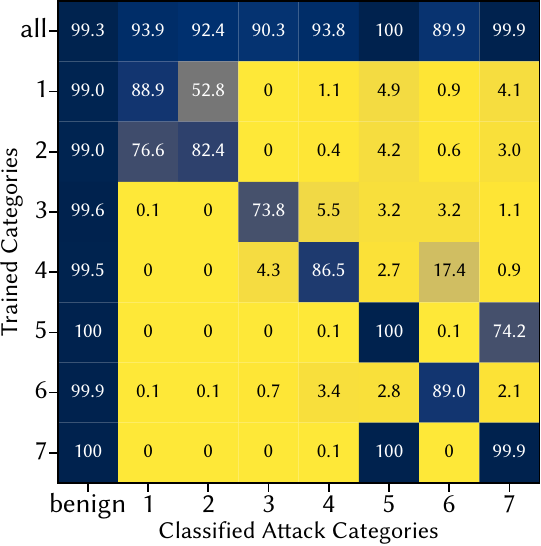}}\quad
  \subcaptionbox{SVM classifier\label{fig:evaluation:single:categories-svm}}[.32\textwidth][c]{%
    \includegraphics[width=.32\textwidth]{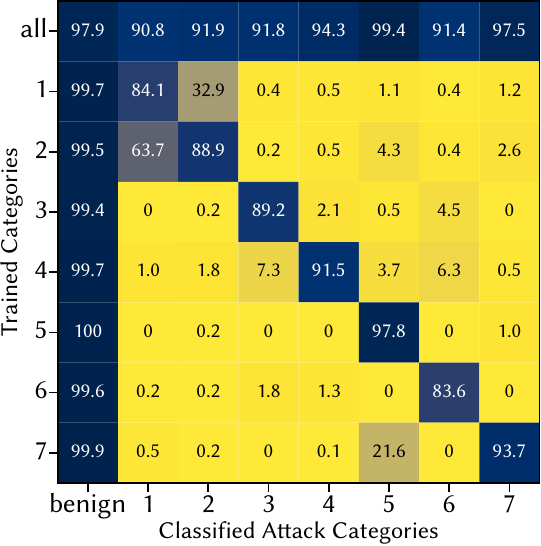}}

  \includegraphics[width=.9\textwidth]{plotting/recall/colorbar}
  \vspace{-0.5em}

  \subcaptionbox{RF classifier\label{fig:evaluation:single:attacks-rf}}[.32\textwidth][c]{%
    \includegraphics[width=.32\textwidth]{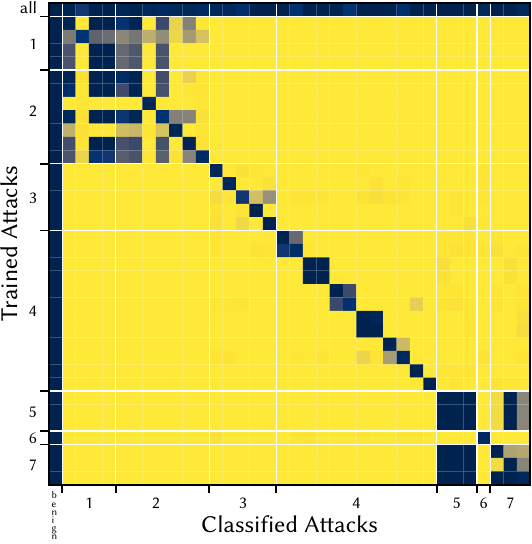}}\quad
  \subcaptionbox{BLSTM classifier\label{fig:evaluation:single:attacks-blstm}}[.32\textwidth][c]{%
    \includegraphics[width=.32\textwidth]{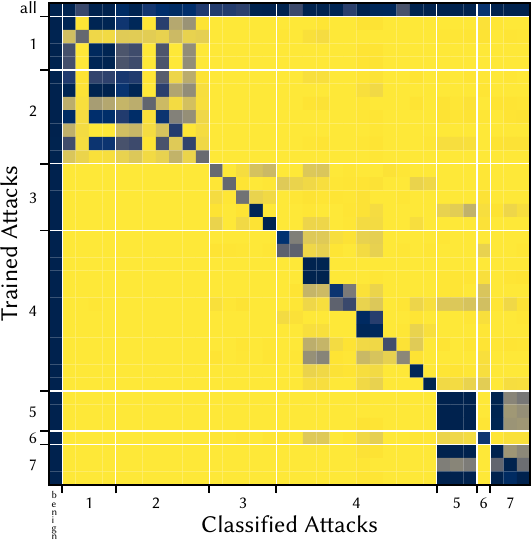}}\quad
  \subcaptionbox{SVM classifier\label{fig:evaluation:single:attacks-svm}}[.32\textwidth][c]{%
    \includegraphics[width=.32\textwidth]{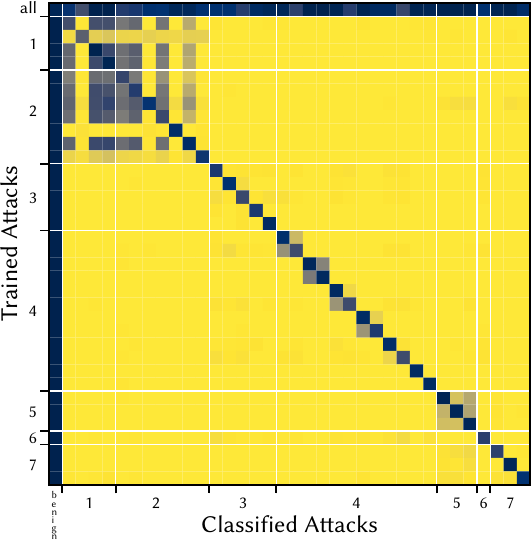}}
  \vspace{-.75em}
  \caption{%
    We specifically trained the classifiers only for single attacks or single attack categories, respectively, using our evaluation methodology (Section~\ref{sec:design:singleattacks}).
    The resulting recall [\%] is visualized as a heatmap similar to Figure~\ref{fig:evaluation:unknown} and underlines that the examined classifiers are hardly able to detect attacks apart from those attacks they were explicitly trained on.
  }
  \Description{%
    Apart from the baseline itself and the column representing benign data, most of the heatmap fields apart from the main diagonal show very low or no recall.
    Fields on the main diagonal achieve around the same recall as the baseline.
    A detailed description and analysis is available in the text.
  }
  \label{fig:evaluation:single}
  \vspace{-.8em}
\end{figure*}

\subsubsection{Omitting Individual Attack Types}

To investigate the classifiers' ability to generalize \emph{within} categories, \ie across attacks that are allegedly much more similar, we also analyze the recall when specific attacks are omitted from training one by one.
We include the corresponding heatmaps in Figures~\ref{fig:evaluation:unknown:attacks-rf}--\ref{fig:evaluation:unknown:attacks-svm}.
The labels on both axes correspond to the categories the individual attacks belong to.

Similar to omitting entire attack categories, we do not observe any significant adverse effects on other attack types when removing individual attack types from the train set, as the effects are again confined to the main diagonal.
Merely for the BLSTM classifier, the attacks from Categories~2 and~3 exhibit a marginally visible effect outside the main diagonal.
These effects indicate an existing, though minimal and insignificant, ability of the BLSTM classifier to abstract attack patterns across attack categories.

Focusing again on the main diagonal, we observe similar patterns across the different classifiers.
For Category~1, the classifiers' ability to detect attacks is primarily unaffected.
However, Attack~1.2, \ie the second attack from Category~1, is an exception, with recall values dropping significantly across all three classifiers.
The dataset describes Attacks~1.1--1.3 as ``Random Value Attacks'' on the pressure measurement without further differentiation between those attacks~\cite{Morrisetal2015Industrial}.
In light of this description, our reported numbers raise the question to which extent Attack~1.2 differs from the others to warrant the observed recall drop.
A manual investigation of the dataset revealed that Attack~1.2 sends pressure values mainly within the normal bounds of the system, \eg{} \num{7.52}, while Attacks~1.1 and~1.3 send values clearly out of bounds, such as \num{0} or \num{6.9e31}.
Thus, we assume that the classifiers cannot detect those attacks within the normal operating bounds without explicit training.

Similar patterns apply to Categories~2,~3,~4, and~6, \ie some unlearned attacks are detected as a variation of another attack independently of the classifier, while others are not identified as such despite originating from the same category.
Yet, Category~7 shows major differences across our evaluated classifiers.
While the RF classifier exhibits no drop in recall, the others (BLSTM and SVM) achieve a reduced recall for Attacks~7.2 and~7.3, motivating an in-depth analysis of the attacks in Category~7.
Attack~7.1 (Device Scan Attack) generates packets whose ``address'' field deviates from the regular address 4, \eg{} setting it to \num{0} or \num{9}, to scan for the presence of those addresses in the network.
Attacks~7.2 and~7.3, in contrast, do not change the address field but introduce ``novel'' Modbus function codes instead.
All classifiers are able to detect the abnormal address, while only the RF classifier is able to detect the malicious function codes without prior training.
This detail shows that classifiers are able to learn the system's normal behavior to some extent and thus detect anomalous deviations from it.
However, this ability is restricted to a specific scenario, and real-world deployments cannot generally trust ML-based IIDSs to detect previously unseen attacks.

Overall, within Categories~1,~2,~4,~5, and~7, most attacks show only minor drops in the recall value, contrary to Categories~3 and~6, with more significant drops across all attacks.
These numbers match our observations when omitting entire categories except for Category~4, which resulted in a very low recall when omitted completely.
It seems that removing some attacks from this category has no significant, immediate effect, while completely removing them drops the recall significantly.
This aspect indicates some form of generalization within Category~4, but not beyond it and to other categories.

\textbf{Takeaways.}
Our analysis highlights the limited generalization capabilities of the analyzed classifiers \emph{within} and \emph{across} attack categories, as many attacks cannot be detected reliably without explicit training.
Further, we observe significant variances in recall rates of previously unseen attacks.
These variances can be (partly) attributed to the similarities between related attacks, but what is interpreted as similar differs between classifiers and does not necessarily match human-created attack categorizations, \ie{} existing dataset labels.
Consequently, despite the predictable nature of ICSs, the analyzed ML-based IIDSs were unable to detect truly novel attacks reliably and thus fail in serving as reliable anomaly detectors.

\subsection{Understanding (Dis-)Similarities of Attacks}
\label{sec:evaluation:singleattacks}

After investigating and validating that ML-based IIDSs hardly generalize to unseen attacks, we now intend to provide a better understanding of the relations between different attacks.
To this end, we explicitly train the classifiers on a single attack (category) only.
By doing so, we can examine the classifiers' ability to detect different attacks beyond the one presented during training.
For the same reasons as in Section~\ref{sec:evaluation:unknownattacks}, we again focus on recall as the primary performance metric and visualize our measured results in Figure~\ref{fig:evaluation:single}.

\subsubsection{Training Specific Attack Categories}

We begin by training on individual attack categories.
We provide the corresponding results in Figures~\ref{fig:evaluation:single:categories-rf}--\ref{fig:evaluation:single:categories-svm}.
Contrary to the previous experiment, we expect the recall values on the main diagonal to be especially high as those fields correspond to explicitly trained attacks.
However, assuming that the classifiers embody anomaly detectors, we would expect noticeable recall apart from the main diagonal.
This observation would indicate some form of general understanding of malicious activities as well as the ability to detect variations of known attacks.

Evidently, detecting novel attack patterns and anomalies between categories is limited to very few cases, with the plots turning out mostly yellow, constituting low recall.
Nevertheless, we observe two main instances of substantial positive recall apart from the main diagonal.
First, we notice an interrelation between Categories~1 and~2:
If one of the two categories is trained, the classifiers also achieve a high recall in the other category, \eg{} when training the BLSTM classifier using attacks from Category~2, attacks from Category~1 are detected with a high recall of \SI{76.6}{\percent}.
As discussed in Section~\ref{sec:evaluation:unknownattacks}, those categories contain attacks that manipulate the pressure reading, which could be the cause of this interrelation.

In addition, Categories~5 and~7 are connected:
If one of the two categories is trained for, the recall in the other category is generally high (up to \SI{100}{\percent}).
According to the dataset's description, one similarity between the six attacks in those categories is that they all employ Modbus function codes that are otherwise not used.
The diagnostic function \texttt{0x08}, for example, is only used in the attacks of Category~5 and the ``Device Scan Attack'' in Category~7~\cite{Morrisetal2015Industrial}.
Thus, we assume that the classifiers learn that only specific function codes are used during normal operation when training on one of the two categories, and, therefore also detect attacks in the respective other category.
Here, the behavior of the SVM classifier is in stark contrast to the others', as it does not derive the same strong relationship between both categories.
Thus, the SVM seems to differentiate benign and malicious traffic on different properties.

Moreover, we observe further differences between the classifiers.
For all but the RF classifier, the recall values of the trained attack categories drop in comparison to the baseline when all attacks are trained, indicating that these classifiers benefit from a general understanding of malicious activities.
The recall values offside the diagonal further underline this observation as the classifiers are partly able to correctly detect attacks they have not been trained on (predominantly noticeable
in the yellow regions of Figures~\ref{fig:evaluation:single:categories-blstm} and~\ref{fig:evaluation:single:categories-svm}).
In contrast, the RF classifier achieves lower recall values outside of the main points of interest but is, therefore, able to detect attacks more reliably when explicitly trained for them.
These observations suggest that the RF classifier realizes a much more targeted training of the attacks it knows from the train set.

Finally, another notable observation is the increase in recall for benign traffic when training on single attacks.
For the RF classifier, the recall reaches \SI{100}{\percent} when single attack categories are trained compared to \SI{99.8}{\percent} before.
A similar improvement can be seen for the SVM classifier.
As the amount of benign traffic in a real-world deployment is expected to be orders of magnitude larger than malicious traffic, this improvement implies a significant decrease in false-positive alarms.
In contrast, both recall and precision, which we calculated additionally, drop when using the BLSTM classifier and training the IIDS exclusively on Categories~1 and~2.

\subsubsection{Training Individual Attacks Exclusively}

In the following, we take a look at the attack detection when the classifier is trained on individual attacks only (Figures~\ref{fig:evaluation:single:attacks-rf}--\ref{fig:evaluation:single:attacks-svm}).
The main observation is that the classifiers generalize within and across categories to some extent, and the SVM classifier generalizes less than RF and BLSTM.
The distinct patterns that we observe do, however, highlight the importance of digging deeper into the classifiers' results to get an accurate impression of their capabilities in a real deployment.

When focusing on Categories~1 and~2, we observe a generalization \emph{within} and \emph{across} the categories for all three classifiers, \eg{} most attacks are detected with a significant recall if the classifiers are trained on Attack~2.4.
Thus, the classifiers are able to generalize well enough to detect the different attacks in those categories, even with such a reduced sample set.
Noticeable exceptions are Attacks~1.2,~2.3,~2.5 for SVM only, and~2.7, which are hardly generalized when training on the other attacks from this attack category.

The pattern for Attack~1.2 is particularly interesting, as it is not correctly detected when training on other attacks in its category.
In contrast, when training the RF classifier on this category, it also detects other attacks in both categories with a significant recall.
As discussed in Section~\ref{sec:evaluation:unknownattacks}, this attack mainly generates malicious pressure readings within the normal operating bounds of the system, while Attacks~1.1 and~1.3 send out-of-bounds readings.
A plausible explanation for this behavior is that abstracting to detect malicious readings \emph{within} bounds enabled the classifier to also detect malicious readings \emph{outside} the normal bounds, but not \textit{vice versa}.
We observe similar generalization patterns between and across attacks of Categories~5 and~7 for RF and BLSTM.
These results again underline that the SVM classifier seems to base its model on different characteristics than the others, at least for this subset of attacks.

Finally, we discuss the pattern that emerges in Category~4 where pairs of two neighboring attacks display an interrelation, \eg{} Attacks~4.1 and~4.2, while interrelations to any other attacks in the category are mostly missing.
Attacks~4.1 and~4.2 both cover ``Setpoint Attacks''~\cite{Morrisetal2015Industrial} and differently manipulate the same communicated value.
Due to the focus on one value, the classifiers can more easily correlate both of these attacks, even if only one of the attacks has been known before.
Similar patterns are also found when looking at the other interrelated attacks from Category~4, \eg{} Attacks~4.4 and~4.5.
Thus, while attacks on the same parameter can be interrelated by the classifiers to some extent, manipulations of other parameters cannot be handled in the same way.

These results further strengthen the assumptions that ML-based IIDSs do not base their detections on abstracted process knowledge but rather on pre-trained signatures of attacks.
Within Category~3, we virtually do not notice any generalization.
Together with the limited generalization in Category~4, where only pairs of attacks are interrelated, we again conclude that human-made categorizations of the dataset, while sensible to humans, are of limited relevance.

\textbf{Takeaways.}
Our second experiment shows that cross-learning \emph{within} categories is limited to a few cases, particularly those patterns where very similar attacks manipulate the same process values.
Thus, ML-based IIDSs, despite contrary promises, only achieve limited generalizability and act much more like signature-based IIDSs.
We further observe major differences between the classifiers' abilities that warrant such in-depth analyses.
Otherwise, we cannot truly understand which attacks an IIDS is able to detect reliably.

\subsection{Implications for IIDS Generalizability}
\label{sec:evaluation:combination}

As the last part of our evaluation, we want to take a holistic look at what can be concluded from the detection capabilities of our examined ML-based IIDSs when combining both of our previous experiments.
We observed a distinct connection between Attack Categories~1,~2,~5, and~7 that performed well when being omitted (\cf Section~\ref{sec:evaluation:unknownattacks}) and the interrelations between Categories~1 and~2 as well as~5 and~7 when being trained individually (\cf Section~\ref{sec:evaluation:singleattacks}).
In the case of RF, we observe a recall of \SI{55.0}{\percent} in Category~2 when trained on Category~1, which matches closely with the \SI{55.8}{\percent} when Category~2 is omitted.
Furthermore, we can make identical observations for Categories~5 and~7 across all evaluated classifiers.

A similar picture emerges when looking at the fine-grained results, i.e., individual attacks:
Those attacks that showed interrelations when classifiers were trained exclusively on them, \eg{} Attack~4.2, show decent recall even when being omitted.
Meanwhile, those attacks that demonstrate hardly any interrelations, \eg{} Attack~4.11, cannot be detected without training for them explicitly.
We also observe that the same attacks are outliers in both experiments, \ie{} Attacks~1.2,~2.3,~2.5 for SVM, and~2.7.
When omitted, those attacks faced a significant drop in recall while other attacks in their category fared much better.
Simultaneously, when training exclusively for those attacks, other attacks were rarely detected.

To conclude, it seems that the observed ability to generalize when omitting attacks can be mainly explained by the interrelations found when training on single attacks.
This finding also puts the observations from Figures~\ref{fig:evaluation:unknown:attacks-rf}--\ref{fig:evaluation:unknown:attacks-svm}, which hinted at slight generalizability, at least within categories, into a better perspective.
Overall, we emphasize that the evaluated ML-based IIDSs do not actually learn normal system behavior but rather directly learn signatures of the attacks that they have been trained on.
Nonetheless, we believe that ML-based IIDSs can constitute an effective and reliable defense mechanism for ICSs.
However, they are in need of additional research and analyses to foster an understanding of their actual capabilities and to allow for accurate assessments and understandings of their benefits outside of (artificial) lab environments.

\section{A False Sense of Security}
\label{sec:discussion}

Applying our evaluation methodology to quantify the generalizability of ML-based IIDSs to unseen attacks (\cf{} Section~\ref{sec:design}), we find that all three of our analyzed classifiers are largely unable to successfully detect unknown attacks (\cf{} Section~\ref{sec:evaluation}), despite scoring high in widely-used performance metrics.
Thus, the results reported by related work can lead to a false sense of security for practitioners.

While we observe cases in which unknown attacks are detected, they mostly result from an overlap of specific attack patterns with trained attacks, rather than the classifiers being able to abstract from known attacks.
Thus, overall, our results support concerns and made claims formulated in literature that ML-based IDSs can only detect variations of known attacks (\cf{} Section~\ref{sec:relatedwork}).
Thereby, we confirm that our considered classifiers only learn signatures of attacks instead of realizing proper anomaly detection.
Consequently, they fail to reap the benefits anomaly-based detection can provide, highlighting that further attention concerning this issue is crucial.

We further observed that the human-made categorization of ``similar'' attacks (within ICS datasets) does not constitute a constructive approach as classifiers tend to source relevant information for the classification from other characteristics than humans.
As a result, research must reconsider its understanding of similarities within the scope of IIDSs.
Otherwise, further research on the generalizability of attack categories could be unnecessarily impaired.

Additionally, our analysis shows that the ability to detect unknown attacks varies between the considered classifiers, highlighting the need to compare multiple classifiers and approaches regarding their abilities in real-world deployments.
The currently used (traditional) evaluation methodology, mainly focusing on recall, precision, or $F_1$-score metrics on a randomly chosen test set, is insufficient when targeting industrial settings with the need for anomaly detection.
Thus, it is incapable of providing researchers and practitioners with a realistic understanding of the capabilities of an IIDS (particularly \wrt{} the protection against novel (unseen) attacks), which is technically the primary goal of any evaluation.

This situation limits the assessment and comparability of different approaches and hinders further advances in this field.
Our proposed methodology addresses this issue by explicitly testing the system on selectively filtered datasets, enabling an in-depth analysis and explanation of the results.
Still, conducting such detailed analyses requires access to a dataset with explicit labeling of (similar) attacks and enough repetitions or observations of each of them.
Unfortunately, the availability of suitable ICS datasets for IDS use is limited~\cite{Bhamareetal2020Cybersecurity,Creechetal2013Generation, Pennekampetal2021Collaboration}.

Given that such analyses are lengthy and require detailed manual analyses of the results, the used dataset, and the aggregated, achieved performance, developing real-world-practical IIDSs is far from trivial.
To at least improve the comparability of different approaches, we argue that developing a precise, comprehensible metric (\eg based on our evaluation methodology) should be a primary concern of future work.
With such a metric at hand, research can then properly compare the IIDSs' abilities to detect unknown attacks, \ie to truly perform anomaly detection.
Regardless, even with corresponding advances in the area of generalizability, the real-world feasibility of ML-based IIDSs is still in its infancy as significant deployment challenges, such as model tuning, sampling rates, and operational changes of the monitored ICSs, await~\cite{Ahmedetal2020Challenges}.

\section{Conclusion}
\label{sec:conclusion}

The convergence of ICSs with the Internet leads to an increasing number of cyberattacks against such systems~\cite{Pennekampetal2019Towards,Serroretal2021Challenges}.
To detect and prevent these attacks, anomaly-based intrusion detection is especially interesting as its detection rate benefits from repetitive communication patterns that frequently occur in industrial settings.
While research focuses on the usage of machine learning enabling industrial IDSs to automatically determine what constitutes benign and malicious behavior, it still remains unclear whether these IDSs have any ability to detect novel attacks as they are typically trained not only on benign behavior but also on attacks~\cite{Junejoetal2016Behaviour-Based}.
Notably, promised detection rates of up to \SI{99}{\percent} are reached by training the IDSs on attack types that are used for training \textit{and} evaluation.

In this paper, we showed that these standard evaluation methods disguise the missing ability of ML-based IIDSs to detect formerly unseen attacks.
More specifically, we proposed a methodology to analyze the ability of ML-based IIDSs to detect novel forms of attacks and applied it to three IIDSs~\cite{Lopez-Perezetal2018Machine}.
We discovered that these IIDSs are widely unable to detect unseen attacks and find detection rates dropping to between \SI{3.2}{\percent} and \SI{14.7}{\percent} for specific unseen attack types.
Furthermore, we proved that the ML-based IIDSs mainly learn specific attack signatures instead of process-specific properties.
Hence, in scenarios where the training data does not cover all possible attacks, the IIDSs can only detect types of attacks that are known beforehand and fail to generalize to new attacks.

We suggest that our methodology should be performed on more IIDSs to ensure comparability \wrt{} the achieved level of generalization and to prevent the manifestation of a false sense of security based on the good performance numbers achieved with state-of-the-art evaluation methods.

\begin{acks}
Funded by the Deutsche Forschungsgemeinschaft (DFG, German Research Foundation) under Germany's Excellence Strategy -- EXC-2023 Internet of Production -- 390621612.
\end{acks}

\bibliographystyle{ACM-Reference-Format-limit}
\bibliography{paper}

\end{document}